\documentclass[12pt]{article}
 \usepackage{epsfig}
 \def\be{\begin{equation}}
 \def\ee{\end{equation}}
 \def\bea{\begin{eqnarray}}
 \def\eea{\end{eqnarray}}
 \usepackage{graphicx}

 \catcode`\@=11
 \def\lsim{\mathrel{\mathpalette\@versim<}}
 \def\gsim{\mathrel{\mathpalette\@versim>}}
 \def\@versim#1#2{\vcenter{\offinterlineskip
 \ialign{$\m@th#1\hfil##\hfil$\crcr#2\crcr\sim\crcr } }}
 \catcode`\@=12

 \catcode`@=12
\begin{document}
 \thispagestyle{empty}
 \begin{flushright}
 UCRHEP-T594\\
 Oct 2018\
 \end{flushright}
 \vspace{0.6in}
 \begin{center}
 {\LARGE \bf $SO(10) \to SU(5) \times U(1)_\chi$ as\\ 
the Origin of Dark Matter\\}
 \vspace{1.2in}
 {\bf Ernest Ma\\}
 \vspace{0.2in}

{\sl Physics and Astronomy Department,\\ 
University of California, Riverside, California 92521, USA\\}
\vspace{0.1in}
{\sl Jockey Club Institute for Advanced Study,\\ 
Hong Kong University of Science and Technology, Hong Kong, China\\} 
\end{center}
 \vspace{1.2in}

\begin{abstract}\
In the decomposition of $SO(10)$ grand unification to 
$SU(5) \times U(1)_\chi$, 
two desirable features are obtained with the addition of one colored 
fermion octet $\Omega$, one electroweak fermion triplet $\Sigma$ 
and one complex scalar triplet $S$ to the particle 
content of the standard model with two Higgs doublets.  They are (1) 
gauge coupling unification of $SU(3)_C \times SU(2)_L \times U(1)_Y$ 
to $SU(5)$, and (2) the automatic (predestined) emergence of 
dark matter, i.e. $\Omega$, $\Sigma$ and $S$, with dark parity 
given by $(-1)^{Q_\chi + 2j}$. It suggests that $U(1)_\chi$ may well be 
the underlying symmetry of the dark sector.
\end{abstract}

 \newpage
 \baselineskip 24pt

\noindent \underline{\it Introduction}~:~
The origin of dark matter and the symmetry which maintains it are important 
issues in particle and astroparticle physics.  A prevalent supposition is 
supersymmetry which admits superparticles as belonging to the dark sector 
if $R$ parity is imposed.  Great hope was attached to the Large Hadron 
Collider (LHC) in discovering supersymmetry at the present 13 TeV 
total center-of-mass energy, but no sign of such has yet been 
reported.  Is there another underlying framework for dark matter? 
More importantly, does this underlying framework provide as well the 
dark symmetry required~\cite{cfs06,m18}, instead of having it imposed 
as in supersymmetry? The answer, as suggested in this paper, is $U(1)_\chi$ 
which is the possible residual symmetry in the breaking of $SO(10)$ to 
$SU(5)$. 

In most studies of $SO(10)$ grand unification, the breaking to a left-right 
extension of the standard model is considered.  In that case, since 
$U(1)_{B-L}$ is an Abelian gauge factor in the decomposition 
$SO(10) \to SU(3)_C \times SU(2)_L \times SU(2)_R \times U(1)_{B-L}$ with the 
electric charge $Q = I_{3L} + I_{3R} + (B-L)/2$, the parity $(-1)^{3(B-L)+2j}$ 
may be used to distinguish matter from dark matter and coincides with 
the $R$ parity of supersymmetry.  Many 
studies~\cite{m92,amrs98,abmrs01,kkr10,kkr09,fh10,mnoqz15,noz15,alrz16,bkn16} 
have been made using this obvious connection. On the other hand, the 
(trivial) breaking of $SO(10)$ to $SU(5)$ is usually considered to be 
uninteresting, because it reduces to studying $SU(5)$ grand unification 
by itself.  The possible residual $U(1)_\chi$ symmetry is treated as an 
unimportant peripheral issue, 
allowing LHC data to put a limit on the $Z_\chi$ boson mass of about 
4.1 TeV~\cite{atlas-chi-17,cms-chi-18}.

Whereas examples of automatic (predestined) dark matter are possible in 
the standard model (SM) itself~\cite{cfs06} or some of its simple gauge 
extensions~\cite{m18}, their origin is unexplained. 
In the context of $U(1)_\chi$, because all SM fermions have 
odd charge $Q_\chi$ and all SM bosons have even $Q_\chi$, the dark sector 
in this framework consists simply of all fermions with even $Q_\chi$ and 
scalars with odd $Q_\chi$.  The stability symmetry of dark matter is thus 
revealed to be $(-1)^{Q_\chi + 2j}$.  Note the important fact that $Q_\chi$ 
is not part of the electric charge, whereas $B-L$ is.  This means 
that $Z_\chi$ is independent of the photon as well as the $Z$ boson, whereas 
the $B-L$ gauge boson would not be.  Note that each complete fermion 
multiplet of $SU(5)$, i.e. $5^*$ or 10, has its own unique $Q_\chi$, 
i.e. 3 or $-1$, whereas the complete fermion multiplet $\underline{16}$ 
of $SO(10)$ has different $B-L$ for its various components, separated by 
its $SU(3)_C \times SU(2)_L \times SU(2)_R$ content. Consequently, 
$Q_\chi$ is more desirable than $B-L$ as a marker of dark matter.  
There is also an important difference in their corresponding phenomenology. 
The $SU(3)_C \times SU(2)_L \times U(1)_Y \times U(1)_\chi$ model assumes 
that there is no intermediate symmetry breaking scale for $SU(5)$, whereas the 
$SU(3)_C \times SU(2)_L \times SU(2)_R \times U(1)_{B-L}$ model assumes 
an intermediate scale where $SU(2)_R \times U(1)_{B-L}$ breaks to $U(1)_Y$. 
Even though $U(1)_\chi \times U(1)_Y$ is equivalent to 
$U(1)_{B-L} \times I_{3R}$, the charged gauge bosons $W_R^\pm$ appear in 
the latter scenario but not in the former.

To find a marker for $SO(10)$ multiplets, consider the decomposition 
$E_6 \to SO(10) \times U(1)_\psi$, with 
\begin{equation}
\underline{27} = (16,-1) + (10,2) + (1,-4).
\end{equation}
In that case, $Q_\psi$ takes the role of $Q_\chi$, 
and $Z_\psi$ is independent of the three neutral gauge bosons 
of $SU(2)_L \times SU(2)_R \times U(1)_{B-L}$.  

Furthermore, under $SU(3)_C \times SU(2)_L \times U(1)_Y \times U(1)_\chi$, 
if the fermions $\Omega \sim (8,1,0,0)$ and $\Sigma \sim (1,3,0,0)$ are 
added together with the scalar $S \sim (1,3,0,-5)$ at the TeV scale, then 
$SU(5)$ gauge unification is achieved with $M_U \sim 10^{16}$ GeV. 
This is a new realization where the particles added to those of the SM 
are all in the dark sector.  It suggests that matter and dark matter 
are related to each other~\cite{m13}.  It is also qualitatively different 
from previous studies requiring $SO(10)$ gauge unification.  In particular, 
the scalar triplet $S$ from the \underline{144} of $SO(10)$ is unique to 
this proposal.

\newpage
\noindent \underline{\it Decomposition of $SO(10)$ to 
$SU(5) \times U(1)_\chi$}~:~
Consider first the \underline{16} representation of $SO(10)$ which contains 
all the SM fermions.  Under its $SU(5) \times U(1)_\chi$ decomposition, it 
is well-known that
\begin{equation}
\underline{16} = (5^*,3) + (10,-1) + (1,-5),
\end{equation}
whereas the \underline{10} representation contains the necessary Higgs 
doublets, i.e.
\begin{equation}
\underline{10} = (5^*,-2) + (5,2).
\end{equation}
Under $SU(3)_C \times SU(2)_L \times U(1)_Y \times U(1)_\chi$,
\begin{eqnarray}
&& (5^*,3) = d^c~[3^*,1,1/3,3] + (\nu,e)~[1,2,-1/2,3], ~~~ (1,-5) = 
\nu^c~[1,1,0,-5], \\ 
&& (10,-1) = u^c~[3^*,1,-2/3,-1] + (u,d)~[3,2,1/6,-1] + e^c~[1,1,1,-1], \\ 
&& \Phi_1 = (\phi_1^0,\phi_1^-)~[1,2,-1/2,-2], ~~~ 
\Phi_2 = (\phi_2^+,\phi_2^0)~[1,2,1/2,2].
\end{eqnarray}
Hence the allowed Yukawa couplings are
\begin{eqnarray}
d^c(u \phi_1^- - d \phi_1^0), ~~~ u^c(u \phi_2^0 - d \phi_2^+), ~~~ 
e^c(\nu \phi_1^- - e \phi_1^0), ~~~ \nu^c(\nu \phi_2^0 - e \phi_2^+),
\end{eqnarray}
as desired.  Note that $U(1)_\chi$ is broken by 2 units as $\phi^0_{1,2}$ 
acquire nonzero vacuum expectation values.  Now the \underline{126} 
representation of $SO(10)$ contains a singlet $\zeta \sim (1,-10)$ 
under $SU(5) \times U(1)_\chi$.  
Such a scalar may be used to break $U(1)_\chi$ at the TeV 
scale and would allow $\nu^c$ (the right-handed neutrino) to obtain a large 
Majorana mass, thereby triggering the canonical seesaw mechanism for 
small Majorana neutrino masses.  This is usually described as 
lepton number $L$ breaking to lepton parity $(-1)^L$~\cite{m15}, but 
here it is clear that it has to do with the breaking of gauge $U(1)_\chi$ 
to $(-1)^{Q_\chi}$.

At this stage, all SM fermions are odd and all SM bosons are even under 
$(-1)^{Q_\chi}$.  Hence they are all even under
\begin{equation}
R_\chi = (-1)^{Q_\chi + 2j}.
\end{equation}
This suggests strongly that a dark sector exists where $Q_\chi$ is even for 
fermions and odd for scalars so they all have odd $R_\chi$.  The next step 
is to identify possible candidates which will also enforce gauge $SU(5)$ 
unification, thus justifying the role of $U(1)_\chi$ as the residual 
symmetry from the breaking of $SO(10)$ to $SU(5)$.

\noindent 
\underline{\it Gauge $SU(5)$ Unification from the Addition of Dark Matter}~:~
To break $SO(10)$ to $SU(5) \times U(1)_\chi$, the scalar \underline{45} 
is used.  Since
\begin{equation}
\underline{45} = (24,0) + (10,4) + (10^*,-4) + (1,0)
\end{equation}
under $SU(5) \times U(1)_\chi$, a nonzero vacuum expectation value (VEV) 
of the (1,0) component will work.  Under 
$SU(3)_C \times SU(2)_L \times U(1)_Y \times U(1)_\chi$,
\begin{equation}
(24,0) = (8,1,0,0) + (1,3,0,0) + (3,2,1/6,0) + (3^*,2,-1/6,0) + (1,1,0,0).
\end{equation}
Hence a nonzero VEV of the (1,1,0,0) component will break 
$SU(5) \times U(1)_\chi$ to the SM gauge symmetry without breaking 
$U(1)_\chi$.  The electroweak symmetry breaking occurs through the 
nonzero VEVs of $\phi^0_{1,2}$ and $U(1)_\chi$ is broken at the TeV scale 
through the scalar $\zeta \sim (1,-10)$ singlet, as discussed in the 
previous section.

The particle content so far consists of all the SM fermions and gauge 
bosons together with two Higgs doublets and one singlet.  There is also 
the $Z_\chi$ gauge boson at the TeV scale.  Experimentally, the three 
gauge couplings corresponding to $SU(3)_C \times SU(2)_L \times U(1)_Y$ 
are measured, but it is well-known that they do not extrapolate to a 
single value at a possible unification scale, based on this particle 
content.  On the other hand, this may be achieved simply with the 
addition of two fermion and one scalar multiplets, all belonging to 
the dark sector at the TeV scale.

Consider the one-loop renormalization-group equations governing the 
evolution of gauge couplings with mass scale:
\begin{equation}
{1 \over \alpha_i(M_1)} - {1 \over \alpha_i(M_2)} = {b_i \over 2\pi} 
\ln {M_2 \over M_1},
\end{equation}
where $\alpha_i = g_i^2/4\pi$ and the numbers $b_i$ are determined by the 
particle content of the model between $M_1$ and $M_2$.  In the SM with one 
Higgs scalar doublet, these are given by
\begin{eqnarray}
SU(3)_C: && b_C = -11 + (4/3)N_F = -7, \\ 
SU(2)_L: && b_L = -22/3 + (4/3)N_F + 1/6 = -19/6, \\ 
U(1)_Y: && b_Y = (4/3)N_F + 1/10 = 41/10,
\end{eqnarray}
where $N_F=3$ is the number of quark and lepton families and $b_Y$ has been 
normalized by the well-known factor of 3/5.  A second Higgs doublet at 
$M_\phi$ would contribute $\Delta b_L = 1/6$ and $\Delta b_Y = 1/10$.

Suppose a colored fermion octet $\Omega \sim (8,1,0,0)$ is added with 
$M_\Omega$ as well as an electroweak fermion triplet $\Sigma \sim (1,3,0,0)$ 
with $M_\Sigma$, both coming from the (24,0) of Eq.~(10).  These are then 
augmented by an electroweak scalar triplet $S \sim (1,3,0,-5)$ with $M_S$, 
coming from the $SO(10)$ scalar representation \underline{144}, i.e.
\begin{equation}
\underline{144} = (5^*,3) + (5,7) + (10,-1) + (15,-1) + (24,-5) + (40,-1) 
+ (45^*,3),
\end{equation}
which contains $(24,-5)$ and thus $(1,3,0,-5)$.  Note that
\begin{equation}
\underline{16}^* \times \underline{10} = \underline{16} + \underline{144}.
\end{equation}
As a result, $\Omega$ contributes $\Delta b_C = (2/3)3 = 2$, $\Sigma$ 
contributes $\Delta b_L = (2/3)2 = 4/3$, and $S$ contributes 
$\Delta b_L = (1/3)2 = 2/3$.  Note that $\Omega$, $\Sigma$, and $S$ all 
have odd $R_\chi$.   Assuming unification at $M_U$ and normalizing 
$\alpha_Y$ by 5/3, the evolution equations are then given by
\begin{eqnarray}
{1 \over \alpha_U} &=& {1 \over \alpha_C} + {5 \over 2\pi} \ln {M_U \over M_Z} 
+ {2\over 2\pi} \ln {M_\Omega \over M_Z}, \\ 
{1 \over \alpha_U} &=& {1 \over \alpha_L} + {1 \over 2\pi} \ln {M_U \over M_Z} 
+ {1 \over 2\pi} \left( {1 \over 6} \right) \ln {M_\phi \over M_Z}  
+ {1 \over 2\pi} \left( {4 \over 3} \right) \ln {M_\Sigma \over M_Z}  
+ {1 \over 2\pi} \left( {2 \over 3} \right) \ln {M_S \over M_Z}, \\ 
{1 \over \alpha_U} &=& {3 \over 5\alpha_Y} - {1 \over 2\pi} \left( 
{21 \over 5} \right) \ln {M_U \over M_Z} 
+ {1 \over 2\pi} \left( {1 \over 10} \right) \ln {M_\phi \over M_Z}, 
\end{eqnarray}
where $\alpha_C,\alpha_L,\alpha_Y$ are evaluated at $M_Z$.  Their central 
values are~\cite{pdg2018}
\begin{equation}
\alpha_C = 0.118, ~~~ \alpha_L = (\sqrt{2}/\pi)G_F M_W^2 = 0.0340, 
~~ \alpha_Y = \alpha_L \tan^2 \theta_W = 0.0102.
\end{equation}

The idea that octets and triplets are important in $SU(5)$ gauge unification 
is not new~\cite{bm84,k93,gl03,m05}.  However, the role of $U(1)_\chi$ 
was not recognized.  Otherwise, the choice here follows closely that of 
Ref.~\cite{m05}.  Note that the chosen particle content is free of 
gauge anomalies even with the inclusion of $U(1)_\chi$.  
Eliminating $\alpha_U$ and using Eq.~(20), the two 
conditions on $M_U$, $M_\Omega$, $M_\Sigma$, $M_S$ and $M_\phi$ are
\begin{eqnarray}
35.538 &=& \ln {M_U \over M_Z} + 0.2564 \ln {M_\Sigma \over M_Z} + 
0.1282 \ln {M_S \over M_Z} + 0.0128 \ln {M_\phi \over M_Z}, \\ 
32.888 &=& \ln {M_U \over M_Z} + 0.5 \ln {M_\Omega \over M_Z} - 
0.3333 \ln {M_\Sigma \over M_Z} - 0.1667 \ln {M_S \over M_Z} - 
0.0417 \ln {M_\phi \over M_Z}. 
\end{eqnarray}
Subtracting the two equations to eliminate $M_U$, the condition 
\begin{equation}
2.650 = 0.5897 \ln {M_\Sigma \over M_Z} + 0.2949 \ln {M_S \over M_Z} + 
0.0545 \ln {M_\phi \over M_Z} - 0.5 \ln {M_\Omega \over M_Z}
\end{equation}
is obtained.  Assuming the neutral component $\Sigma^0$ to be dark matter, 
it has been shown some years ago~\cite{ms09} that $M_\Sigma \simeq 2.3$ TeV. 
Using that value and assuming the second Higgs doublet to have 
$M_\phi \simeq 500$ GeV, the constraint
\begin{equation}
0.654 = 0.2949 \ln {M_S \over M_Z} - 0.5 \ln {M_\Omega \over M_Z}
\end{equation}
may be satisfied for example with $M_S = 100$ TeV and $M_\Omega = 1.53$ TeV, 
in which case $M_U = 4.32 \times 10^{16}$ GeV and $\alpha_U = 0.0276$.

\noindent \underline{\it Phenomenology of the Dark Sector}~:~
The dark sector consists of the scalar $S \sim (1,3,0,-5)$ and the fermions 
$\Sigma \sim (1,3,0,0)$, $\Omega \sim (8,1,0,0)$.  They are the only particles 
beyond those of the SM (with two Higgs doublets) in addition to $Z_\chi$ 
and the scalar singlet $\zeta \sim (1,1,0,-10)$ which breaks $U(1)_\chi$.  
Consider first 
the colored fermion octet $\Omega$.  It is just like the gluino of 
supersymmetry, except that it is stable here because there are no scalar 
quarks.  However, because it has 
strong interactions, bound states do exist~\cite{kk84,gh85,ck05} from the 
exchange of gluons.  These gluinonia would then decay into quark pairs. 
At the LHC, they may be searched for, as shown in 
Ref.~\cite{ck05}.  

If $M_S > M_\Sigma$, then $S$ decays to $\nu \Sigma$ through the 
$U(1)_\chi$ allowed $f_S S^* {\nu}^c \Sigma$ Yukawa coupling and the 
neutrino $\nu-\nu^c$ mixing $\sim \sqrt{m_\nu/M_R}$, with a decay rate
\begin{equation}
\Gamma = {f_S^2 m_\nu M_S \over 16 \pi M_R} \left( 1 - {M_\Sigma^2 \over 
M_S^2} \right)^2.
\end{equation}
Assuming $m_\nu = 0.1$ eV, $M_R = 4$ TeV, $f_S = 0.01$, $M_S = 100$ TeV, 
$M_\Sigma = 2.3$ TeV, then this decay lifetime is $1.3 \times 10^{-10}s$ 
which is certainly acceptable cosmologically.

In this scenario, $\Sigma^0$ is stable.  Its relevance as dark matter 
has been studied in detail~\cite{ms09}.   In particular, the radiative 
splitting of $\Sigma^\pm$ with $\Sigma^0$ is known~\cite{s95} to be 
positive from gauge boson interactions, but is limited~\cite{cfs06} to 
167 MeV.  This means that whereas $\Sigma^0$ is the dark matter today, 
its relic abundance is determined in the early Universe with the 
coexistence of $\Sigma^\pm$, i.e. all annihilation channels involving 
$\Sigma^0,\Sigma^\pm$ have to be considered.  This was done in 
Ref.~\cite{ms09} and
\begin{equation}
2.28 < M_\Sigma < 2.42~{\rm TeV}
\end{equation}
was obtained.  Similar results are obtained in supersymmetry with a pure 
wino as dark matter~\cite{betal16}.  The difference is that the wino has 
many other interactions which are absent in the case of $\Sigma$. 
As for direct detection, $\Sigma^0$ does not couple to the $Z_\chi$ or 
$Z$ or any scalar at tree level, but may interact with quarks in one and 
two loops.  However, these effects are known to be small~\cite{hin10}, 
with the spin-independent cross section below $2 \times 10^{-47}cm^2$.

If $M_S < M_\Sigma$, then $S^0$ is dark matter.  Since it is a scalar, 
it has possible quartic interactions with the two Higgs doublets and 
one singlet.  Any 
such interaction must be suppressed to avoid the constraint from 
direct-search experiments because all $R_\chi$ even scalars couple to 
quarks.  This means that the annihilation cross section 
of $S^0$ to scalars would be too small, so its relic abundance should 
again be determined by gauge interactions, as in the case 
of $\Sigma$.

\noindent \underline{\it Phenomenology of the $U(1)_\chi$ Sector}~:~
The contribution of the SM particles to $b_\chi$ is
\begin{equation}
b_\chi = {1 \over 40} \left( {2 \over 3} \right) [5(3)^2 + 10(-1)^2] N_F 
+ {1 \over 40} \left( {1 \over 3} \right) [2(2)^2] = {169 \over 60},
\end{equation}
where the factor 1/40 has been inserted to normalize $Q_\chi$. 
The addition of a second Higgs doublet at $M_\phi$ contributes 
$\Delta b_\chi = 1/15$, and those of $\nu^c$ and the scalar singlet 
$\zeta \sim (1,1,0,-10)$ contribute $\Delta b_\chi = 25/12$, whereas $S$ 
contributes $\Delta b_\chi = 5/8$.  Hence
\begin{equation}
{1 \over \alpha_U} = {1 \over \alpha_\chi} - {1 \over 2\pi} \left( 
{671 \over 120} \right) \ln {M_U \over M_Z} + {1 \over 2\pi} \left( 
{1 \over 15} \right) 
\ln {M_\phi \over M_Z} + {1 \over 2\pi} \left( {5 \over 8} \right) \ln 
{M_S \over M_Z} + {1 \over 2\pi} \left( {25 \over 12} \right) \ln 
{M_R \over M_Z}.
\end{equation}
\begin{figure}[htb]
\vspace*{0.5cm}
\hspace*{1cm}
\includegraphics[scale=0.8]{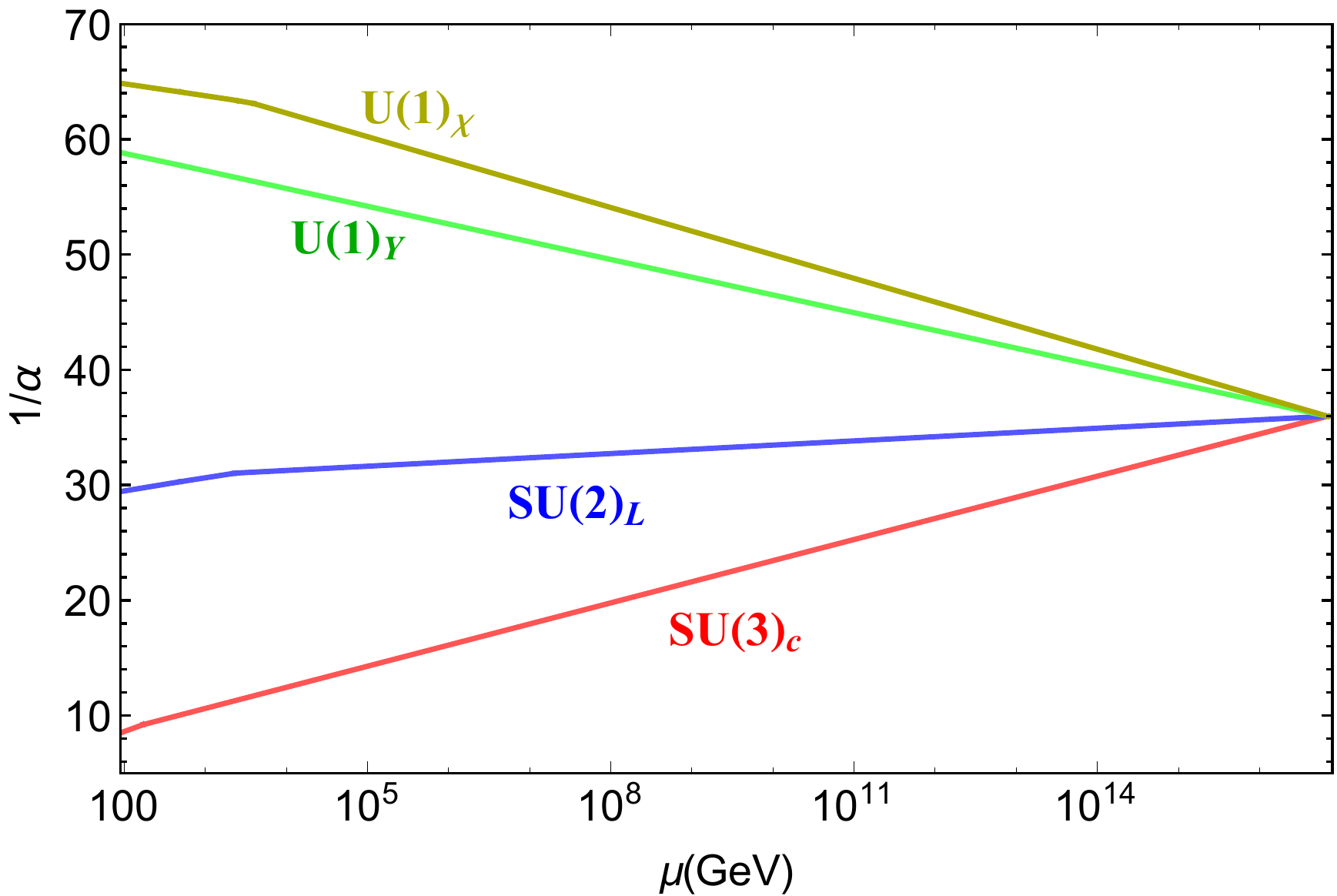}
\caption{Running of $1/\alpha_i$ with energy scale.}
\end{figure}
Using the previously determined values, $\alpha_\chi = 0.0155$ at $M_Z$ 
is obtained.  The one-loop evolutions of 
$1/\alpha_\chi$, $3/5\alpha_Y$, $1/\alpha_L$, and $1/\alpha_C$ are depicted 
as functions of energy scale in Fig.~1.

The symmetry breaking of $SU(2)_L \times U(1)_Y \times U(1)_\chi$ occurs 
through the VEVs $v_{\chi,1,2}$ of $\zeta \sim (0,0,-10)$, 
$\phi_1^0 \sim (1/2,-1/2,-2)$, and $\phi_2^0 \sim (-1/2,1/2,2)$, 
where the values of $(I_{3L},Y,Q_\chi)$ for each are shown. 
As a result, the mass-squared matrix spanning $(Z,Z_\chi)$ is given by
\begin{equation}
{\cal M}^2_{Z,Z_\chi} = \pmatrix{(g_Z^2/2)(v_1^2+v_2^2) & -(g_Z g_\chi/\sqrt{10})
(v_1^2+v_2^2) \cr -(g_Z g_\chi/\sqrt{10})(v_1^2+v_2^2) & 5g_\chi^2 v_\chi^2 + 
(g_\chi^2/5)(v_1^2+v_2^2)}.
\end{equation}
Using $M_{Z_\chi} = 4.1$ TeV from the LHC lower bound, the $Z-Z_\chi$ mixing 
is then at most
\begin{equation}
\theta_{Z-Z_\chi} \simeq \sqrt{2 \over 5} {g_\chi \over g_Z} \left( {M_Z \over 
M_{Z_\chi}} \right)^2 = 1.85 \times 10^{-4},
\end{equation}
which is consistent with precision electroweak measurements~\cite{pdg2018}.

\noindent \underline{\it Two Variations with Dirac Neutrinos}~:~
Instead of the canonical scenario with Majorana neutrinos from the seesaw 
mechanism, the $U(1)_\chi$ extension allows for two interesting variations 
with Dirac neutrinos.

(A) Replace the scalar $\zeta \sim (1,-10)$ from the \underline{126} 
of $SO(10)$ by the scalar $\zeta' \sim (1,-5)$ from the \underline{16}. 
Now $\langle \zeta' \rangle \neq 0$ breaks $U(1)_\chi$ but 
$\nu^c \sim (1,-5)$ cannot obtain a Majorana mass.  In fact, $\zeta'$ always 
appears together with $(\zeta')^*$.  In other words, because 
of the chosen particle content, $\zeta'$ does not interact singly with any 
combination of the available fields.  After spontaneous 
symmetry breaking, the resulting Higgs scalar 
$H_\chi = \sqrt{2} (Re(\zeta')-v_\chi)$ behaves as a particle with even 
$R_\chi$ even though its original $Q_\chi$ is odd.

In this scenario, both baryon number $B$ and lepton number $L$ are 
conserved, with $\Omega$ and $\Sigma$ having $B=L=0$, and $S$ having 
$B=0$ and $L=-1$.  The Yukawa term $f_S S^* {\nu}^c \Sigma$ discussed 
earlier forms the link between them and the Dirac neutrinos.   
Again $\Sigma^0$ may be chosen as stable dark matter, 
because there can be no lighter collection of 
particles with an odd number of fermions carrying $B=L=0$. 
Similarly, if $S^0$ is lighter than $\Sigma^0$, then it is stable because 
there can be no lighter collection of particles with an even number of 
fermions carrying $B=0$ and $L=-1$.  The origin of dark matter is 
again $U(1)_\chi$ which allows $B$ and $L$ to be generalized to include 
dark matter.

(B) On top of $\zeta \sim (1,-10)$ from the \underline{126} of $SO(10)$, 
add the scalar $\zeta'' \sim (1,15)$ from the \underline{672}.  Let the 
$U(1)_\chi$ symmetry be broken by the latter and not the former, i.e. 
$\langle \zeta'' \rangle = v_\chi$, but $\langle \zeta \rangle = 0$. 
In that case, there is again no Majorana mass term for $\nu^c$, and neutrinos 
are Dirac fermions.  However, there are now the allowed terms
\begin{equation}
\zeta^* \nu^c \nu^c, ~~~ \zeta^* S S.
\end{equation}
They imply that $\zeta$ has $L=-2$ and $S$ has $L=-1$. 
In other words, the proposal of Ref.~\cite{m18-1} of leptonic dark matter 
$S^0$ with scalar dilepton mediator $\zeta$ is realized, where $\zeta$ 
decays only to two neutrinos.  Assumming $\zeta$ to be light (10 to 100 MeV), 
the self-interacting dark matter $S^0$ may then explain~\cite{kkpy17} the 
central flatness of the density profile of dwarf galaxies~\cite{detal09}.  
On the other hand, $\zeta$ has a large production cross section through 
Sommerfeld enhancement at late times.  Its decay to electrons and photons 
would disrupt~\cite{gibm09} the cosmic microwave background (CMB) and be 
ruled out~\cite{bksw17} by the precision observation 
data now available~\cite{planck16}.  Here $\zeta$ decays only to neutrinos, 
thus solving this important problem.  Note that if $S$ is not a triplet 
but a singlet, then the Yukawa terms $\zeta'' SSS$ and $\zeta'' \zeta S$ 
would be allowed~\cite{m18-1}, in which 
case $U(1)_L$ breaks to $Z_3$ and $S^*$ transforms as $\zeta$, so it cannot 
be dark matter.  As it is, the fact that $S \sim (1,3,0,-5)$ allows it 
to be stable dark matter, as well as a contributer to the gauge unification 
of $SU(5)$ from $SU(3)_C \times SU(2)_L \times U(1)_Y$.

\noindent \underline{\it Concluding Remarks}~:~
It has been proposed in this paper that matter and dark matter are 
unified under $SO(10)$ 
which breaks to $SU(5) \times U(1)_\chi$ at $M_U \sim 7 \times 10^{16}$ GeV. 
Matter consists of fermions with odd $Q_\chi$ charge under $U(1)_\chi$ and 
bosons with even $Q_\chi$.  It encompasses all particles of the SM (extended 
to include two Higgs doublets) as well as the $U(1)_\chi$ gauge boson $Z_\chi$ 
and the corresponding Higgs singlet which breaks $U(1)_\chi$.  Dark matter 
consists of fermions with even $Q_\chi$ and scalars with odd $Q_\chi$. 
To achieve $SU(5)$ gauge unification from 
$SU(3)_C \times SU(2)_L \times U(1)_Y$, they are chosen to be a colored 
fermion octet $\Omega \sim (8,1,0,0)$, an electroweak fermion triplet 
$\Sigma \sim (1,3,0,0)$, and a complex electroweak scalar triplet 
$S \sim (1,3,0,-5)$ at or below the TeV scale.  The dark parity 
$R_\chi = (-1)^{Q_\chi + 2j}$ is identified as the stabilizing symmetry 
for dark matter.  Either $\Sigma^0$ or $S^0$ (whichever is lighter) 
is a good dark-matter candidate.

In the canonical scenario, neutrinos are Majorana with $U(1)_\chi$ broken by 
the scalar singlet $\zeta \sim (1,-10)$ under $SU(5) \times U(1)_\chi$. 
However, if $\zeta' \sim (1,-5)$ or $\zeta'' \sim (1,15)$ is used, 
neutrinos could be Dirac, and in the latter case, $\zeta$ itself may be 
the light scalar dilepton mediator for the self-interacting dark matter 
$S^0$.  Since $\zeta$ decays only to two neutrinos, it does not disrupt 
the CMB, unlike other models where the mediator decays to electrons and 
photons.

To verify this proposal that $U(1)_\chi$ is the origin of dark matter, 
the production of $Z_\chi$ would be a necessary first step.  However, 
its mass is not precisely predicted, only that it should be at the TeV scale. 
At present, the LHC bound~\cite{atlas-chi-17,cms-chi-18} is about 4.1 TeV.  
However, a more detailed study~\cite{kkm18} shows that it can be improved.
Other new particles to look for are the bound states of $\Omega$, i.e. 
gluinonia, which are possible~\cite{ck05} with higher luminosity 
at the LHC.

It should also be pointed out that if supersymmetry is imposed 
on $SO(10) \to SU(5) \times U(1)_\chi$, then the origin of $R$ parity is 
again traced to $Q_\chi$.  In other words, there is no need to impose it 
to distinguish the would-be identical Higgs and lepton superfields in 
the minimal supersymmetric standard model, because they now have different 
$Q_\chi$.  Hence it could turn out that supersymmetry is there after all, 
but to explain dark matter, $U(1)_\chi$ is still the key.  In that case, 
gauge coupling unification comes about from the presence of the gluino 
(acting as $\Omega$), the wino (acting as $\Sigma$), the bino, and two 
higgsino doublets (replacing $S$), as well as complete multiplets of 
squarks and sleptons.

As remarked earlier, there are three equivalent markers of dark matter: 
$(-1)^{3(B-L)+2j}$, $(-1)^{Q_\chi+2j}$, and $(-1)^{Q_\psi+2j}$.  They are all even 
for the known SM particles and odd for would-be particles of the dark 
sector.  The choice of the latter is somewhat arbitrary and many studies 
have been made.  In Ref.~\cite{m92}, $(-1)^{3(B-L)}$ was considered in the 
context of supersymmetry and 
$SO(10) \to SU(3)_C \times SU(2)_L \times SU(2)_R \times U(1)_{B-L}$ was 
advocated but not $SO(10) \to SU(5) \times U(1)_\chi$, although $SU(5)$ 
by itself was discussed and discarded.  Subsequent work all follow this 
lead.  In Refs.~\cite{amrs98,abmrs01}, supersymmetric $SO(10)$ was 
considered with conserved $(-1)^{3(B-L)}$.  The dark sector consists of 
all superpartners of the SM particles and the various Higgs multiplets 
used to break the symmetry at various scales.
In Ref.~\cite{kkr10}, the scalar singlet and elctroweak doublet 
contained in the \underline{16} of $SO(10)$ are considered as dark matter 
in a nonsupersymmetric context.  In Ref.~\cite{kkr09}, 
$SO(10) \to SU(5) \times (-1)^{Q_\chi}$ was considered.  This work is closest 
to the present proposal, but there $Z_\chi$ is superheavy and unobservable, 
whereas here the SM particles have $Q_\chi$ charges and couple to $Z_\chi$ 
at the TeV scale.   Also, the chosen dark sector is completely different.
In Ref.~\cite{fh10}, nonsupersymmetric $SO(10)$ was considered using 
$(-1)^{3(B-L)}$ as a marker, but $U(1)_\chi$ was not mentioned and 
$Z_\chi$ is explicitly absent.  However, this model is close to the present 
proposal in its dark-matter content, i.e. the electroweak fermion triplet 
$\Sigma$. In Refs.~\cite{mnoqz15,noz15}, nonsupersymmetric $SO(10)$ was 
considered with various breaking scales.  There is again no $Z_\chi$ and 
the scenarios for dark matter are different.  In Ref.~\cite{alrz16}, 
nonsupersymmetric $SO(10) \to SU(5) \times U(1)_\chi$ was mentioned but 
it breaks to the SM at the unification scale.  Hence $Z_\chi$ is again 
superheavy.  The dark sector mimics those of supersymmetry, i.e. the 
gauginos and the higgsinos.  
In Ref.~\cite{bkn16}, nonsupersymmetric $SO(10)$ was considered, where  
fermions in a vectorlike $\underline{10}_L$ + $\underline{10}_R$ of $SO(10)$ 
belong to the dark sector.  These are analogs of the higgsinos in 
supersymmetry, i.e. the fermionic partners of the scalar \underline{10} 
of $SO(10)$ which contains the Higgs bidoublet.

In summary, these references are related but also very different from 
the scenario discussed in this paper, which is indeed new and not 
contained in any previous study of this specific topic.

\noindent \underline{\it Acknowledgement}~:~
This work was supported in part by the U.~S.~Department of Energy Grant 
No. DE-SC0008541.

\baselineskip 16pt
\bibliographystyle{unsrt}

\end{document}